*Article*

# A new measure of risk using Fourier analysis

Michael Grabinski [1] and Galiya Klinkova [2],[*]

[1]  Neu-Ulm University; Wileystr. 1, 89231 Neu-Ulm, Germany, michael.grabinski@uni-neu-ulm.de
[2]  Neu-Ulm University; Wileystr. 1, 89231 Neu-Ulm, Germany, galiya.klinkova@uni-neu-ulm.de
[*]  Correspondence: michael.grabinski@uni-neu-ulm.de

**Abstract:** We use Fourier analysis to access risk in financial products. With it we analyze price changes of e.g. stocks. Via Fourier analysis we scrutinize quantitatively whether the frequency of change is higher than a change in (conserved) company value would allow. If it is the case, it would be a clear indicator of speculation and with it risk. The entire methods or better its application is fairly new. However, there were severe flaws in previous attempts; making the results (not the method) doubtful. We corrected all these mistakes by e.g. using Fourier transformation instead of discrete Fourier analysis. Our analysis is reliable in the entire frequency band, even for frequency of 1/1d or higher if the prices are noted accordingly. For the stocks scrutinized we found that the price of stocks changes disproportionally within one week which clearly indicates speculation. It would be an interesting extension to apply the method to crypto currencies as these currencies have no conserved value which makes normal considerations of volatility difficult.

## 1. Introduction

Almost all financial products (e.g. stocks) have neither constant prices nor fixed interest. The price of a stock is supposed to reflect the future earnings of the underlying company. As it is impossible to *know* the future, one can only speculate about it. The fluctuating price of a stock can be too high or too low compared to the future to come. In essence there is certain risk involve in buying stocks and the like.

There are many ways to estimate risk involved. One of the most common tools is to calculate the mean quadratic deviation around an average (standard deviation) which leads to the so-called volatility. Here we will investigate in a related but fairly new way. Of course there are more advanced ways to access risk in the financial world rather than just looking for volatility. As an example consider (Fahling et al. 2018).

In section 2.1 we will explain the Fourier analysis. Anything (e.g. prices) which changes over time does so *slowly* or *rapidly*. There is a frequency of change. High frequencies mean rapid changes and low frequencies mean slow ones. This is exactly comparable to a tone of an orchestra. Any sound is a variation in pressure. The different instruments (and even a single instrument) produce pressure changes with higher and lower frequencies. A Fourier analysis will give the amplitudes (strengths) of each frequency of (sinusoidal) pressure change.

Applying this to e.g. stock prices will lead to something like x% of the changes of a stock price was due to daily change (high frequency), y% came from monthly changes (medium frequency), and z% from annual changes (low frequency). Of course a real spectrum will consist of many more frequencies. The maybe earliest attempts to access risk via Fourier can be found in (Bormetti et al. 2010) and (Baruník and Křehlík 2018), respectively. However, these works focus on (valuable) technical aspects and not so much on measures for risk assessment in the world of finance.

(Schädler 2018) used Fourier (probably independent of the above) to find a tool for determining risk in financial products. It has been extended by (Schädler and Steurer 2019) and applied to portfolio selection (Fahling et al. 2019). The general idea was as follows. The value (profitability) of a company may increase (or decrease) because of new products, new technology, new inventions, new markets, and many more factors. All this will take time. To develop a new product may take a year, to introduce it the market another one. Even very quick measures might take several months. A change of price of a company within e.g. a week must be purely speculative and cannot be connected to the (conserved) company value. As an example consider one week in fall 2008 when the stock of Volkswagen AG (a German car maker) gained and lost fourfold within a week (Appel and Grabinski 2011). Nobody visiting the company rather than the stock market would have noticed anything special during that week.

Using Fourier to analyze the stocks will show the frequencies of the changes in stock price. If the spectrum is dominated by high frequencies it means that these changes are speculative. (Schädler 2018) called this irrationality. The general idea is marvelous. However, there are some severe flaws. Especially, frequencies over $1/(10 \text{ days}) = 1/(2 \text{ weeks})$ were neglected because they could not be included for principle (technical) reasons. As it turns out, these high frequencies are essential. Some other flaws such as using the quadratic amplitudes produced results which are not there in reality. For more detail please see section 2.2.

The main purpose of this paper is to fix these problems and draw conclusions from it. Firstly, we transformed the stock prices so that frequencies as high as $1/(1 \text{ day})$ can be reliably used. (As we considered daily prices only, $1/(1 \text{ day})$ is the limit. However, using more frequently listed prices would allow arbitrarily high frequencies) Secondly, we used a Fourier transformation rather than a discrete analysis. For more details please see chapter 3.

In chapter 4 we present our results. An indeed we found a typical "irrationality" for frequencies $\geq 1/(5 \text{ days})$. These results correspond to "gut-feeling" as crazy ups and downs within the stock market appear not rarely within one week and are forgotten in the next one.

We close with conclusions and future work in chapter 5. Though our work is precise, it consumes lots of computing power. So some simplification is desirable. It may also lead to much more useful conclusions from the Fourier transformed price.

It appears to be also very interesting to apply our method to other financial products especially crypto currencies. A risk assessment via volatility is very tricky there, cf. the newer works of (Almeida et al. 2023), (Bowala and Sigh 2022), and (Irfan et al. 2023). Fourier analysis promises new insights.

## 2. General method and previous shortcomings

This section provides an overview how a Fourier transformation can be used to give a risk measure of financial assets. 2.1 provides the general idea including a brief summary of the mathematics behind it.

In 2.2 the general problems are discussed and especially the shortcomings of (Schädler 2018) and (Schädler and Steurer 2019).

*2.1. The general method*

The Fourier transformation is an over 200-year-old tool mostly applied to analyze frequencies in a signal (spectral analysis) or to solve an arbitrary set of linear partial differential equations. First, the Fourier series is introduced.

Any periodic function $f(t)$ can be written as a series of harmonic functions:

$$f(t) = \sum_{k=0}^{\infty} a_k cos(k \cdot \omega t) + b_k sin(k \cdot \omega t) \tag{1}$$

Here a period of $T$ has been assumed so that $\omega = 2\pi/T$. The coefficients $a_k$ and $b_k$ are determined by

$$a_k = \frac{2}{T}\int_0^T dt\, f(t) \cdot cos(k \cdot \omega t) \quad and \quad b_k = \frac{2}{T}\int_0^T dt\, f(t) \cdot sin(k \cdot \omega t) \tag{2}$$

A proof of Eq. (2) is performed by inserting $f(t)$ of Eq. (1) into Eq. (2) and performing the integration. A Fourier series exists only if the integrals of Eq. (2) exist. Of course anything can be found in books like e.g. (Bronshtein et al. 2007)

The interpretation of the Fourier transformation is as follows. The function $f(t)$ is changing over time. If it is changing "rapidly," it is a high frequency, if it is changing "slowly," it is a low frequency. In a general function $f(t)$ there are of course slow and rapid changes. It is a mixture of frequencies. The $a_k$ and $b_k$ are the amplitudes of the set of frequencies. So the Fourier transformation analyses quantitively how much a financial assets changes e.g. on a daily, monthly, or yearly bases.

Before discussing the application for financial assets, there are other "versions" of a Fourier transformation. Instead of having discrete frequencies, continuous frequencies can be applied. $a_k$ or $b_k$ are then becoming a function rather than a set of discrete parameters. This leads to the so-called Fourier transformation. The Fourier transformed $\tilde{f}(\omega)$ of a not necessarily periodic function $f(t)$ is defined by

$$\tilde{f}(\omega) \equiv \int_{-\infty}^{\infty} dt\, f(t) \cdot e^{-i\omega t} \tag{3}$$

As usual $i^2 \equiv -1$. There is also a backward transformation given by

$$f(t) = \frac{1}{2\pi} \int_{-\infty}^{\infty} dt\, \tilde{f}(\omega) \cdot e^{i\omega t} \tag{4}$$

The proof of Eq. (3) or (4) is again performed by inserting Eq. (3) into Eq. (4) or vice versa. The Fourier transformed exists if the integral in Eq. (3) exists. Eq. (4) is the continuous analogue to Eq. (1). The sum in Eq. (1) is transformed into an integral and the discrete coefficients $a_k$ and $b_k$ are now a function $\tilde{f}(\omega)$. The absolute value $|\tilde{f}(\omega)|$ is the "amplitude" of this particular frequency $\omega$. Please do not be confused that $\tilde{f}(\omega)$ has complex values (even if $f(t)$ is real). The identity

$$e^{i\omega t} = cos(\omega t) + i \cdot sin(\omega t) \tag{5}$$

shows that the real part of $\tilde{f}(\omega)$ corresponds to $a_k$ and the imaginary part to $b_k$. In this sense one sometimes speaks of the cosine or sine transformed function. In the same token one may use Eq. (5) to rewrite Eq. (1) into

$$f(t) = \sum_{k=-\infty}^{\infty} c_k \cdot e^{k \cdot i\omega t} \tag{6}$$

with $c_k \in \mathbb{C}$ given by

$$c_k = \frac{1}{T}\int_0^T dt\, f(t) \cdot e^{-k \cdot i\omega t} \tag{7}$$

Eqs. (6) and (7) are only a different writing of Eqs. (1) and (2).

With this short course in mathematics, we can show how this can be used in evaluating financial assets. The price of anything (and especially financial products) can be displayed by a function $f(t)$. If the financial product is a stock or similar, its price may change rapidly or with high frequency. In the classical interpretation of (Fama 1970) there are random fluctuations. Meanwhile it had been proven that they are chaotic (Klinkova and Grabinski 2017b). At least for stocks where there is an underlying value of a company, the change in price should reflect the change in company value or better expected future value.

In a fixed interest financial product there is no risk (except for the underlying currency) and therefore no fluctuation. Therefore fluctuations are taken for a measure of risk. The most often used approach is volatility or standard deviation. There are more advanced methods which are cum grano salis based upon volatility.

Using Fourier analysis to scrutinize risk is a method suggested by (Schädler 2018). The idea behind it goes as follows. The (changing) price of a stock should reflect the (future) value of the company. However, the price of a stock may change within a millisecond (or shorter). The (conserved) value of a company may change over months or even years. Obviously fast changes are due to pure speculation. As a typical example consider the stock of VW, German car manufacturer. In fall 2008 its stocks gained and loosed fourfold within a week (Appel and Grabinski 2011). The explanation for it was simple. There was a takeover poker with Porsche, a German sports car manufacturer. However, the "real" value of Volkswagen did not butch at all during that week. That led to the concept of conserved value (cannot change rapidly) and speculation (Appel and Grabinski 2011). It is also the main idea behind using Fourier analysis to access risk. If the amplitudes for high frequencies (e.g. $|c_{k \gg 1}|$ of Eq. (7)) are "big" compared to the others, it is "irrational" (a term phrased by (Schädler 2018)), speculation and with-it risk is dominant. Therefore (Schädler 2018) introduced the ratio

$$\frac{\sum_{k=0}^{N}|c_k|}{\sum_{k=0}^{\infty}|c_k|} \tag{8}$$

$|c_N|$ is the amplitude of a frequency $N\omega$ which is still considered reasonable (e.g. 1/(3 months)). The closer the ratio of Eq. (8) is to 1, the lesser is speculation or irrationality. To choose $N$ is of course arbitrary. However, one may also scrutinize the entire spectrum of $|c_k|$ and draw conclusions from it.

In Table 1 some results from (Schädler 2018) have been displayed. For exactly how these irrationalities are calculated see (Schädler 2018) and (Schädler and Steurer 2019). Some of it will be discussed in section 2.2. These three particular stocks will be reconsidered in chapter 3:

**Table 1.** Some results taken from (Schädler 2018).

| Company | Irrationality |
|---|---|
| BASF SE | 77.8% |
| SAP SE | 79.2% |
| Deutsche Bank AG | 82.8% |

Up to now we have shown the fairly new method of using Fourier transformation to access risk. In what follows we will discuss problems and shortcomings of this approach.

*2.2. Previous shortcomings*

The general technique from the last section has been used in e.g. physics for centuries for e.g. analyzing radio signals from far away solar systems in order to discover orbiting planets. It is also a standard tool to solve linear differential equations. The theory and especially its applications in the real world are for sure correct.

For the application here there is a severe difference. We do not have a *function $f(t)$*. We just have discrete quotes for prices. This does not look like a severe problem as Figure 1 shows the end of day prices of BASF SE (a German chemical giant) for 5050 trading days.

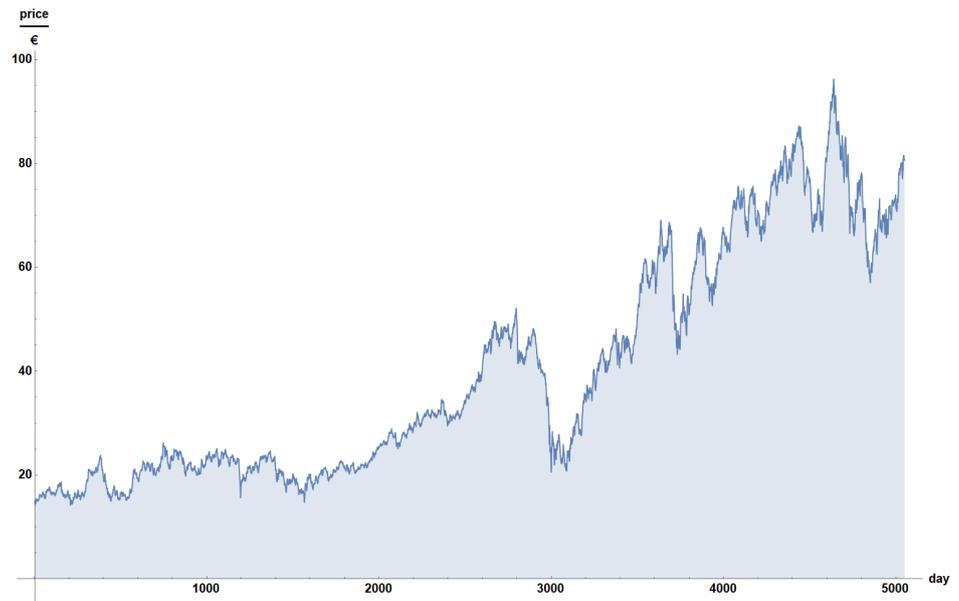

**Figure 1**. Price of BASF stock between January 2, 1997 and December 30, 2016.

Though the graphics of Figure 1 looks like a perfect approximation for a function $f: \mathbb{R}^+ \to \mathbb{R}^+$, it isn't one for principle reasons. There are 5050 discrete points which aren't equally distant. Within 20 years there are typically 5050 trading days as the stock market formally closes over the weekend and some holidays. Formally speaking, any integral over a "function" displayed in Figure 1 is zero.

Stock prices are quoted much more often than daily. Though there is partly a price every millisecond, sometimes it takes many minutes for a new price. But even considering any quoted price, the problem of discrete prices does remain. Furthermore such historic values are hard to get, make different stocks not comparable as their prices are quoted at different times, and would lead to tremendous amounts of data. From chapter 3 it would be clear that even the 5,050 prices considered here do lead a huge CPU time.

In experimental physics (especially astronomy) there are also discrete values which should be Fourier transformed. They are not a discrete series in itself. Normally it was not possible to measure the signals continuously. This is in contrast to the financial data. Prices on the stock market do not *exist* between two quotes. As the price of a stock is in almost all circumstances far away from the conserved value (Appel and Grabinski 2011) of the company considered, it does not make sense to speculate about continuous prices.

A function being Fourier transformed via Eqs. (1) and (2) must be periodic. The transformation involving Eqs. (3) and (4) need an integrable function running form minus infinity to plus infinity. Obviously neither requirement is met by a "function" like in Figure 1. Of course it is easy to make the function periodic just as it is done in solid states physics. Or for using Eqs. (3) and (4) it can be assumed a function running from minus infinity to plus infinity just by setting it to zero outside the regime displayed in Figure 1.

In (Schädler 2018) and (Schädler and Steurer 2019) the problem has been omitted by using a discrete Fourier transformation. It is the analog to a continuous Fourier transformation for a set of discrete numbers. It is textbook knowledge (Bronshtein et al 2007) and has been applied to e.g. random numbers (Lanczos and Gellai 1975) where a "normal" Fourier analysis is not possible as a random function is not integrable (at least not by using Riemann integrals). A discrete Fourier analysis is considered an approximation to the Fourier analysis of Eqs. (1) and (2). It is not clear how big the mistake of this approximation is here but it seems to be small.

The main limitation in (Schädler 2018) and (Schädler and Steurer 2019) was that frequencies $> 1/(10 \text{ days})$ were not considered. It has been because all values were observed on a daily basis only. So frequencies $> 1/(1 \text{ day})$ are for sure nonsense.

Considering frequencies at least ten times lower avoids for sure nuisance effects. On the other hand, 10 days are two trading weeks. What if the main "irrational" fluctuations appear in this period? Just by observing the stock market it looks like that there are typically highly fluctuating weeks rather than months. As an archetype example just consider the "crazy" week of the VW stock in fall 2008 mentioned in section 2.1 (Appel and Grabinski 2011). By considering frequencies $< 1/(10\ days)$ such purely speculative fluctuations are excluded. And indeed, chapter 4 will prove that the main effect appears within frequencies $> 1/(5\ days)$ (and $\leq 1/(1\ days)$).

Besides considering too low frequencies, (Schädler 2018) and (Schädler and Steurer 2019) considered the squares of the amplitudes $|c_k|$. Squares make "small things smaller and big things bigger." However, squaring is unlike the Fourier transformation not a linear transformation. Therefore the results depend on the chosen dimension. Here it depends on whether the stocks are quoted in € or Yen and the time is measured in days or seconds. Taking the square will amplify differences. Why not taking the fourth, sixth, eighth, or tenth power? In doing so differences which are below the measurement accuracy will suddenly appear to be within the accuracy.

It does not help that this squaring is often wrongly used. Even by performing a least square fit it is supposed to be a "least absolute value fit" (Grabinski and Klinkova 2020). In (Schädler 2018) and (Schädler and Steurer 2019) the squaring was used because they wanted to scrutinize the *power spectrum*. However, this is no justification here. In the before mentioned radio signals from outer space a power spectrum or squared amplitudes do make sense. Directly measured is an electric (or magnetic) field. The amplitudes of this electromagnetic field are almost meaningless from the physical point of view. As the energy or here energy current is a conserved quantity which is proportional to the square of an amplitude, scrutinizing the squares of the amplitudes scrutinizes the (conserved) energy pro time which is also known as power (measured in e.g. Watts). (Therefore the name power spectrum)

Transferring this one-to-one into the financial world is ludicrous. The price of a stock and also its square is not a conserved value (Appel and Grabinski 2011). Therefore the fluctuations which are due to speculation and with it risk. It does not help to square the prices or here their changes per time.

## 3. The new method used

Having highlighted the shortcomings in previous publications, it is now easy to fix these problems.

First one has to make a function out of the data points in Figure 1. Financial products with no fluctuations (e.g. fixed interest) are growing or decaying exponentially. As we are not considering any fluctuations between two quoted prices, it is only natural to connect data points by exponential curves just as they were fixed interest assets with no fluctuations. It is straight forward (but maybe puzzling) to do this for all 5,050 prices. To see the difference, we display the first ten trading days of BASF SE in Figure 2. As a next step we deduct the average interest from each exponential line connecting two data points. Without fluctuations we would have a straight line now. As a last step we subtract the value of the first data point (which is equal to the last data point) from each data point. In Figure 3 it is shown schematically how the sperate data points of Figure 1 transforms into a Fourier transformable function. On the r.h.s. of Figure 3 we have now a perfectly fine Fourier transformable function containing the fluctuations only. The same must be done for all data being analyzed.

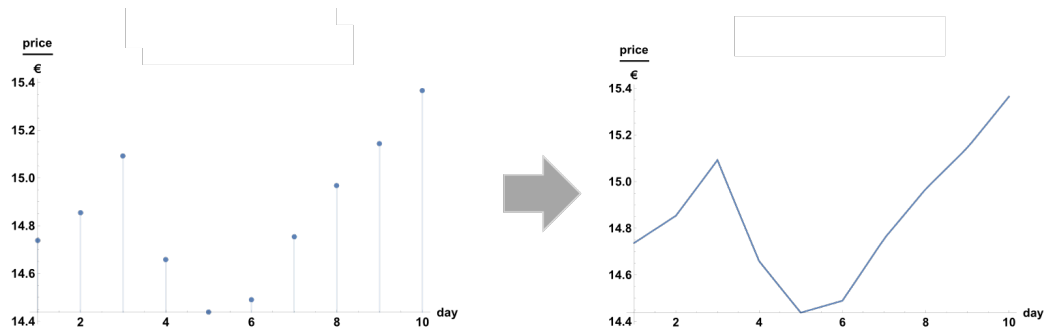

**Figure 2**. Transfomation of separate prices of BASF SE to continuous function.

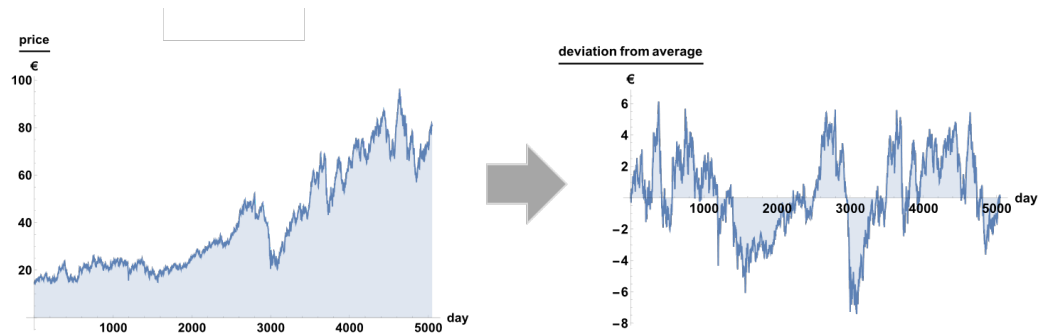

**Figure 3.** Transformation of separate prices of BASF SE into function.

The Fourier transformable function on the r.h.s. of Figure 3 should be transformed by using Eq. (3) rather than (7) because one week would contain only five $c_k$ which must be compared to over 5,000 "ordinary" amplitudes. The function on the r.h.s. of Figure 3 is a piecewise exponential function. Its Fourier transformed can be obtained very easily analytically. Its absolute value is a "simple" but *very* long formula. In Eq. (9) the result has

$$|\tilde{f}(\omega)| = \sqrt{\left(\left(-\frac{5.879096792750167(-\text{Cos}[\omega] + \text{Cos}[5050\omega])}{\omega} + \ldots + 3.558113329776304 \times 10^{-10}\text{Sin}[5051\omega]\right)\right)^2)} \quad (9)$$

been displayed for the example of BASF SE. Displaying the Fourier transformed (or more precisely its absolute value) in Eq. (9) in full (even in this not very nice format) will take over 500 pages. To handle this function is possible with a computer algebra such as Mathematica only. In Figure 4 the function from Eq. (9) has been displayed. It has some simi-larities to the 5,050 amplitudes in (Schädler 2018) (figure 1 there). Please note that Figure 4 contains a function and not separate values. The details can be seen in the enlargement in Figure 5. As we see, frequencies of $1/(\text{several days})$ are slightly more pronounced as the peaks of slightly lower frequencies. At first glance it looks like a nuisance effect. However, it is a true effect as we can easily show if we plot the same as in Figure 5 but for SAP SE. In Figure 6 we see a dramatically enhanced peak. This proves the hint in section 2.2 that the trading within one week does produce most change. Many more things can be seen from the Fourier transformed. The spectrum of BASF shows only a slight increase for high frequencies. Else Figure 4 and Figure 5 show a typical $1/\omega^\alpha$ with $\alpha > 0$ behavior. This is typical for some random or better chaotic fluctuations where lower frequencies are more important. In the case of SAP (Figure 6) one sees no general decay of the amplitudes but a peak for frequencies $1/(\text{several days})$. For very low frequencies (not displayed here) SAP shows a $1/\omega^\alpha$ behavior. This means that in the long run the price of SAP adjusts to the underlying company value. In shorter periods of time SAP takes speculative prices only. This is in perfect accordance to (Appel and Grabinski 2011) who also analyzed the SAP stock by completely other means.

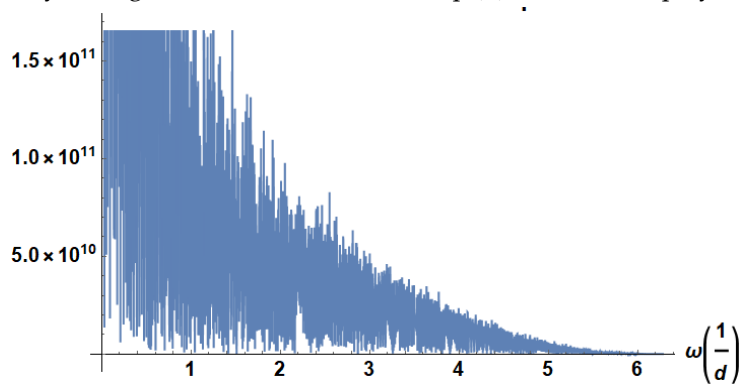

**Figure 4.** Fourier transformed of BASF SE price

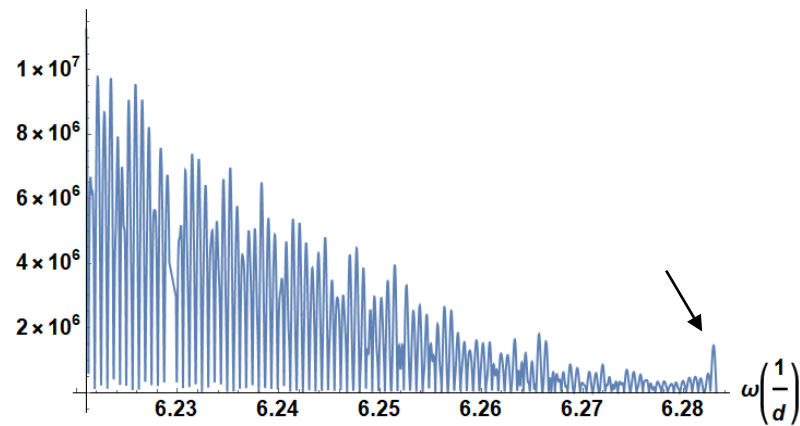

**Figure 5.** $|\tilde{f}(\omega)|$ of BASF SE for frequencies of $1/(50 \text{ days})$ to $1/(1 \text{ day})$

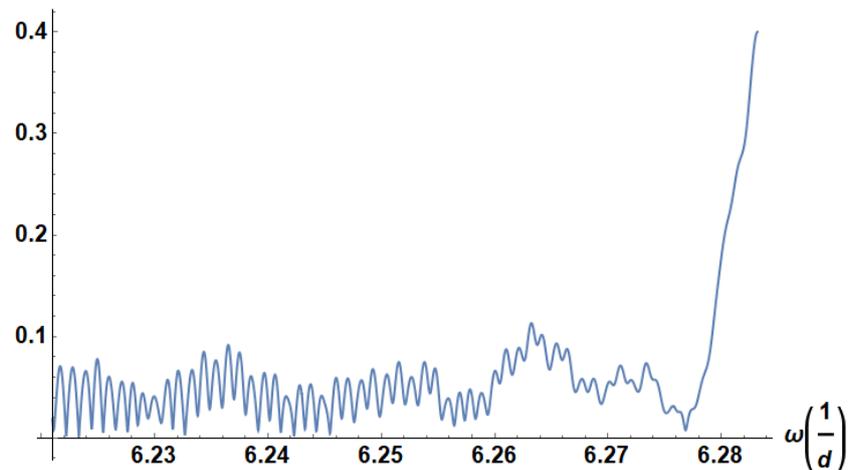

**Figure 6.** $|\tilde{f}(\omega)|$ of SAP SE for frequencies of $1/(50 \text{ days})$ to $1/(1 \text{ day})$

Many more things can be said from the spectra of stock prices. However, they are quite individual. In chapter 4 we will give *one* analysis which may be used for all stocks universally which is in accordance to (Schädler 2018) and (Schädler and Steurer 2019).

Though formulas like in Eq. (9) are quite long, all calculations did not consume much computing power. On an ordinary laptop a CPU time of five to ten minutes will produce Figure 4, Figure 5, or Figure 6 and also the complete formula in Eq. (9). To evaluate the Fourier transformed $|\tilde{f}(\omega)|$ one need an integration instead of summation in Eq. (8). This will be the topic of chapter 4. The principle is simple but the numerical integration will consume lots of CPU time of typically six weeks. Parallelization can reduce it accordingly.

With our means we still needed five days. Therefore we evaluated the three stocks of BASF SE, SAP SE, and Deutsche Bank AG only.

## 4. Quantitative analysis of the Fourier transformed

As the Fourier transformed $|\tilde{f}(\omega)|$ shows the speculative behavior by an undue increase at some frequencies (expected range here $1/(5 \text{ days})$ to $1/(1 \text{ day})$) one should consider the following function

$$F(\omega) = \int_0^{\omega/day} d\omega' |\tilde{f}(\omega')| \qquad (10)$$

in the rage $0 \leq \omega \leq 2\pi$. As $F(\omega)$ is the "sum of the amplitudes" up to $\omega$, an undue increase at some $\omega$ would unveil the frequency where speculation takes place. As the relative rather than absolute amplitudes are essential, one should normalize $F(2\pi)$ to 1. This is easily done by dividing $F(\omega)$ of Eq. (10) by

$$F_{tot} = \int_0^{2\pi/day} d\omega' |\tilde{f}(\omega')| \qquad (11)$$

which will make the results for different stocks comparable. Needless to say that the integrals in Eqs. (10) and (11) cannot be executed analytically. Of course they can be calculated numerically. As $|\tilde{f}(\omega)|$ shows 5,050 oscillation, a sufficient accuracy is given by dividing each oscillation into 1,000 parts. So the numerical integration is essentially performed by inserting 5,050,000 values between 0 and $2\pi/day$ into $|\tilde{f}(\omega)|$. It will lead to 5,050,000 functional values. Adding them up in accordance to the integration limits will solve the integrals numerically. As stated, $|\tilde{f}(\omega)|$ is a simple but very long formula, cf. Eq. (9). Getting the over five million functional values consumes roughly six weeks of CPU time.

The normalized plot of $F(\omega)$ for SAP is displayed in Figure 7. On this scale there is hardly anything unusual. However, displaying frequencies of between $1/(50 \text{ days})$ and

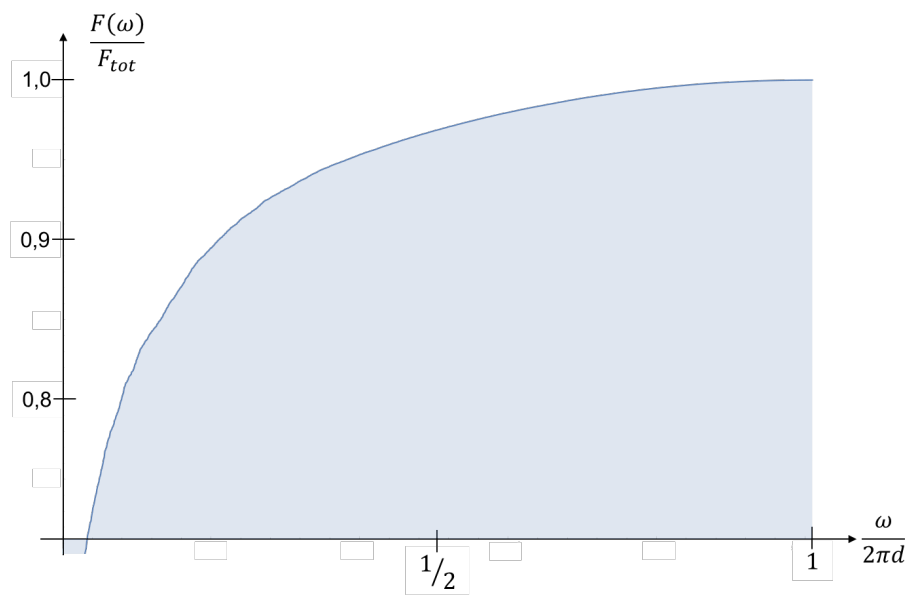

**Figure 7.** Normalized plot of $F(\omega)$ from Eq. (10) for SAP

$1/(1 \text{ day})$, will lead to the graphics of Figure 8. On this scale the summed-up amplitudes are increasing almost linear which is natural as for high frequencies the slope in Figure 7 is fairly low. In the high frequency range of Figure 8, 10 % of the displayed frequency range accumulated to the last 30 % of the functional value of $F(\omega)$.

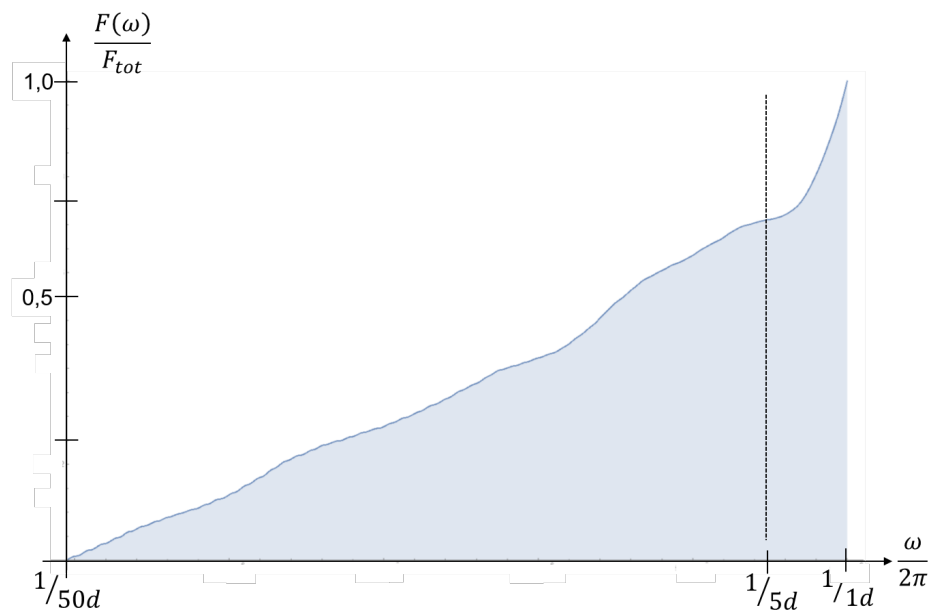

**Figure 8.** Normalized plot of $F(\omega)$ for SAP for frequencies $1/(50\ \text{days})$ to $1/(1\ \text{day})$

The same can be done for Deutsche Bank AG (DB). The result is displayed in Figure

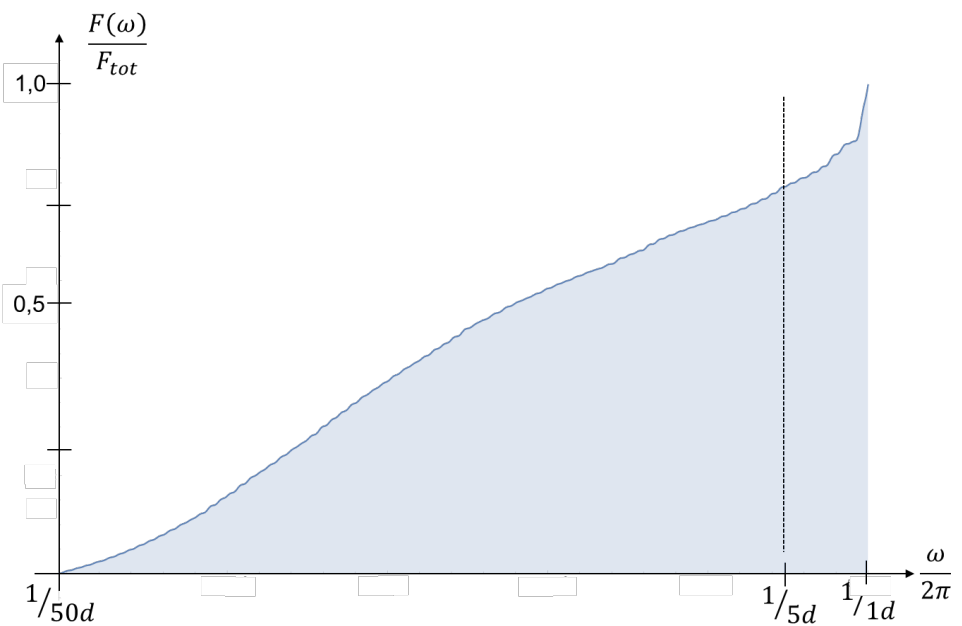

**Figure 9.** Normalized plot of $F(\omega)$ for DB for frequencies $1/(50\ \text{days})$ to $1/(1\ \text{day})$

9 Compared to SAP the effect is much smaller. But in the frequency around $1/(25\ \text{days})$ there is an increase. It is still a frequency range were the conserved value of a company should not change as in five weeks it is hardly possible to change the value of a company. So this is also a hint of speculation but probably not within Deutsche Bank itself. As Deutsche Bank is a lender and investor in many companies like SAP, it is probably a smeared-out effect from underlying companies.

As a last stock we consider BASF. In Figure 10 one can see almost no effect like before. Its looks like in Figure 7 where the entire frequency range has been displayed. This is in accordance with Figure 5 and the comment under it. BASF is a company producing goods which are needed especially for other companies. There is little room for speculation whether its products become into fashion or not.

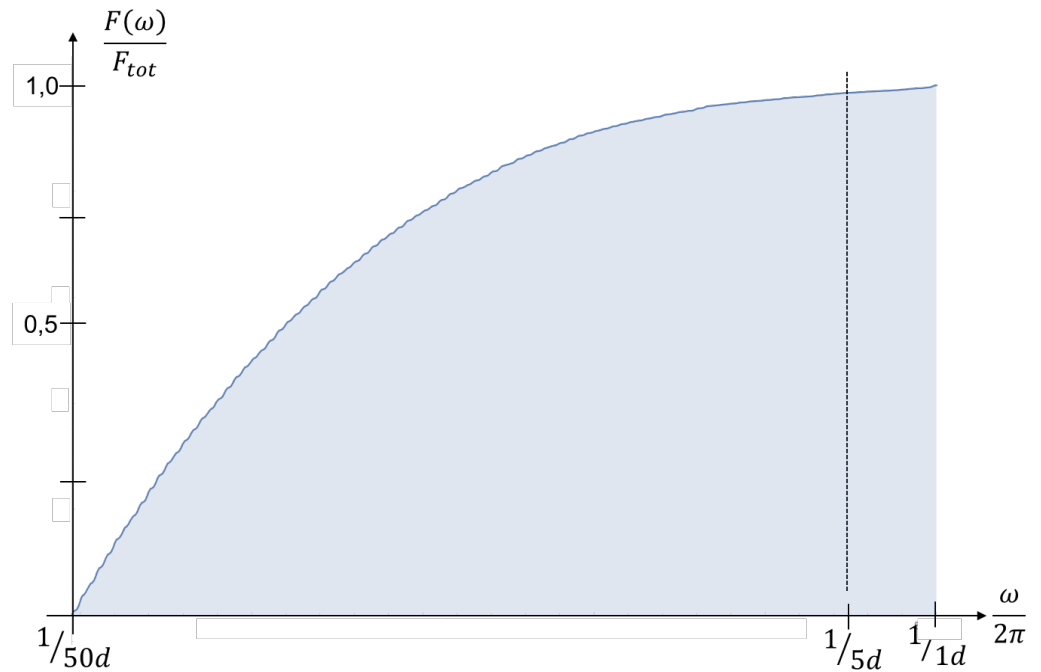

**Figure 10.** Normalized plot of $F(\omega)$ for BASF for frequencies $1/(50 \text{ days})$ to $1/(1 \text{ day})$

Just for completeness, a ten times zoomed scale is displayed in Figure 11. One sees a disproportional increase for frequencies of $1/1(\text{day})$ and slightly below. BASF does show almost no speculation by the analysis of Fourier transformation. That does not mean that the BASF stock is a safe bet. As BASF uses lots of energy like gas and oil as a raw material, its value will increase or decrease with the energy market which is highly volatile. However, BASF is using hedging to be independent of short-term price fluctuations in the energy market. Therefore the Fourier analysis focusing on the frequencies of changes does not show any unusual things.

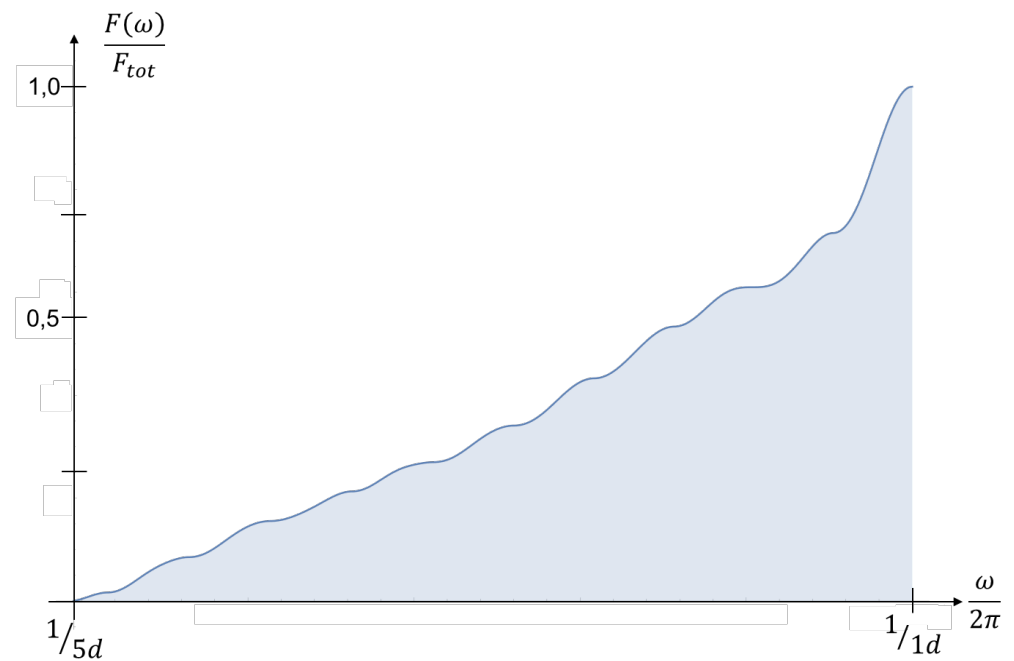

**Figure 11.** Normalized plot of $F(\omega)$ for BASF for frequencies $1/(5 \text{ days})$ to $1/(1 \text{ day})$

Before closing this section we will compare our findings to the ones of (Schädler 2018) and also say a few words about the difference between Fourier analysis and volatility.

For the stocks considered here, (Schädler 2018) found the following results which are summarized in Table 2. We found that SAP is most irrational, BASF least and Deutsche

Table 2. Results from (Schädler 2018)

| Company | Irrationality | Volatility |
|---|---|---|
| BASF SE | 77.8% | 23.7% |
| SAP SE | 79.2% | 32.5% |
| Deutsche Bank AG | 82.8% | 33.1% |

Bank in between. Compared to the results of Table 2, SAP and Deutsche Bank changed places. Furthermore, BASF showed almost no irratonality in our analysis. Please note that this has to do with the method (Schädler 2018) used and its flaws mentioned in section 2.1. The frequency range considered here was not considered. So it is impossible that (Schädler 2018) has found any of our results.

There is also a column about volatility in Table 2. The results for volatility are in accordance with our results from Fourier transformation. Volatility is related to findings by a Fourier transformation but it is far from being identical. Volatility measures changes in price over an entire period. We here focused on short term (high frequency) changes. So the advantage of Fourier transformation is to measure a spectrum of changes. Sometimes volatility is measured on different time scales (monthly, annual,…). This goes to the direction of our analysis. However, volativity must always analyze some time period which has enough results to perform statistics or calculate a meaningful standard deviation. With our analysis the Fourier transformed $\tilde{f}(\omega)$ contains *all* time scales. The limit in our approach is only that the smalles frequenciy is $1/(1 \text{ day})$ as we considered dayly price only. But this is of course not a principle limition. Had we take prices every second, the limit were $1/(1 \text{ second})$

## 5. Conclusions and future work

Classically it is assumed that stocks and the like adjust to their true value by random fluctuation in price (Fama 1970). The height of these fluctuations measures the risk involved as one can never be sure whether the current price is over or under the true value. So it comes natural to measure volatility as it is the average quadratic deviation from the mean.

From (Appel and Grabinski 2011), (Schefczyk 2012), and (Klinkova and Grabinski 2017b) we know that the fluctuations are chaotic rather than random. This makes averages and other statistical operations at least doubtful (Grabinski and Klinkova 2019). Furthermore random fluctuations force a Gaussian distribution. At least from (Grabinski and Klinkova 2020) we know that even the so-called fat tail is due to a wrongly applied statistics. Even speaking of fluctuations (being them random or chaotic) around a *true* value is misleading. We have a conserved value (Klinkova and Grabinski 2017a) plus a speculative part which fluctuates.

We investigated in analyzing (fluctuating) prices of financial products via Fourier transformation. In doing so all the above is not relevant. As the (conserved) value changes slowly only, short term price changes cannot have its origin in a change of conserved value. Fourier transformation measures the amount of change over the frequencies. The price function $f(t)$ is transformed into $\tilde{f}(\omega)$ the so-called Fourier transformed, cf. Eqs. (3) and (4).

We scrutinized just three stocks of SAP, Deutsche Bank, and BASF. We did it to prove the principle and to uncover the flaws of (Schädler 2018). As a next step one should scrutinize the waste amount of stocks (Schädler 2018) and (Schädler and Steurer 2019) analyzed.

There are two other ways to improve this publication. Firstly, one should try to simplify our method as it currently consumes lots of computing power. There appears to be little chance to do this directly. Most likely the very long formula for $|\tilde{f}(\omega)|$ (e.g. Eq. (9)) cannot be simplified. Even a "brute force" attempt to simplify $|\tilde{f}(\omega)|$ failed. We used e.g. "FullSimplify" in Mathematica. We had to abort it after over 50 hours of CPU time and 2.6 TB of RAM in the process. But instead of using $|\tilde{f}(\omega)|$ one may use its real or imaginary part only. It is of course not correct. However, if it would lead to similar results in the three cases presented here, one should scrutinize it further. The formula for the real or imaginary part of $\tilde{f}(\omega)$ is *much* simpler than $|\tilde{f}(\omega)|$ from Eq. (9). It is even possible to integrate it analytically. (The main obstacle will be that the real and imaginary show negative and positive values and its integral is essentially zero. Taking the absolute value is of no help as it would lead to little improvement compared to considering $|\tilde{f}(\omega)|$)

Secondly, we analyzed $|\tilde{f}(\omega)|$ in a not standard way. We showed successfully that it has too high values for high frequencies but we did not have a measure for it. Furthermore, probably much more can be extracted from $|\tilde{f}(\omega)|$. But currently we can only discuss it. So we have no quantitative comparison between different stocks.

It is an interesting question what a Fourier analysis of crypto currencies will show. There is a waste amount of newer publications analyzing risk in cryptocurrencies, e.g. (Almeida et al. 2023), (Bowala and Sigh 2022), and (Irfan et al. 2023). Analyzing cryptocurrencies makes the classical analysis via volatility, etc. at least questionable. The main reason is that cryptocurrencies do not have a conserved value. So there is no adjustment towards it and no fluctuations around it, be them random or chaotic. A Fourier analysis of these prices will reveal a frequency spectrum where no frequency is per se "abnormal." But it may or may not show a "typical" frequency. The result is important to understand *how* people speculate with crypto currencies.

**Author Contributions:** Both authors contributed equally to the entire publication.

**Funding:** There has been no external funding.

**Data Availability Statement:** All relevant data are published within this paper.

**Acknowledgments:** The authors are grateful to Tobias Schädler for sharing the historical data of the stocks considered.

**Conflicts of Interest:** the authors declare no conflicts of interest. or state "The authors declare no conflicts of interest." Authors must identify and declare any